\documentclass[tightenlines,aps,floatfix,preprint,
nofootinbib]{revtex4}

\usepackage{psfig,float}
\usepackage{amsmath,amsfonts,graphicx,bm}
\usepackage{times}

\newcommand{\ep}{\epsilon}
\newcommand{\be}{\begin{equation}}
\newcommand{\ee}{\end{equation}}
\newcommand{\ba}{\begin{eqnarray}}
\newcommand{\ea}{\end{eqnarray}}

\begin{document}


\begin{flushright}
\vbox{
\begin{tabular}{l}
UH-511-1092-06
\end{tabular}
}
\end{flushright}

\title{
Electroweak gauge boson production at hadron colliders through ${\cal O}(\alpha_s^2)$
}

\author{Kirill Melnikov
        \thanks{e-mail: kirill@phys.hawaii.edu}}
\affiliation{Department of Physics and Astronomy,
          University of Hawaii,\\ 2505 Correa Rd. Honolulu, HI 96822}  
\author{Frank Petriello\thanks{frankjp@phys.wisc.edu}}
\affiliation{
Department of Physics, University of Wisconsin, Madison, WI  53706
\medskip
} 

\begin{abstract}
\medskip
We describe a calculation of the ${\cal O}(\alpha_s^2)$ QCD corrections to the 
fully differential cross section for $W$ and $Z$ boson production 
in hadronic collisions. The result is fully realistic in that
it includes spin correlations, finite width effects, 
$\gamma-Z$ interference  and allows for the application of arbitrary cuts on the leptonic decay products of the $W$ and $Z$. 
We have implemented this calculation into a numerical program.  
We demonstrate the use of this code by presenting phenomenological results for several future LHC analyses and recent Tevatron measurements, 
including the $W$ cross section in the forward rapidity region and the central over forward cross section ratio.

\end{abstract}

\maketitle


\section{Introduction}
The study of the electroweak gauge bosons $W$ and $Z$ is an important part of the physics 
programs at the Tevatron and the LHC.  Their large production rates and clean experimental signatures 
facilitate several important measurements, such as the determination of the electroweak parameters 
$M_W$ and ${\rm sin}^2 \theta_W$, and the extraction of the parton distribution functions of the proton~\cite{wmeas}.
The idea to use $W$ and $Z$ production to monitor hadron collider luminosities has begun to be studied 
at the Tevatron and will continue to be investigated at the LHC~\cite{lum}.  
During the initial stages of LHC running, $W$ and $Z$ bosons will be used to calibrate  
lepton energy scales, test the uniformity of the electromagnetic calorimeter and study the tracker alignment~\cite{gianotti}.
Searching for deviations from Standard Model predictions in di-lepton events with 
large invariant mass, missing energy, or transverse momentum probes extensions 
of the Standard Model which contain new gauge bosons or other exotic resonances.

The extensive experimental program described above requires accurate theoretical predictions, and many 
calculations describing electroweak gauge boson production are available.  For example, 
the fully differential level ${\cal O}(\alpha_s)$ next-to-leading order (NLO) QCD \cite{nlo} 
and one-loop electroweak corrections \cite{ew1}
to $W,Z$ production have been known for many years.
The NLO QCD corrections are incorporated into the 
parton-shower event generator MC@NLO~\cite{webber}, which provides a consistent 
description of ${\cal O}(\alpha_s)$ hard emission effects together with a leading-logarithmic resummation 
of soft and collinear QCD radiation. The $p_T$-spectra of 
$W$ and $Z$ bosons are computed with next-to-leading-logarithmic accuracy and are incorporated into the 
program RESBOS~\cite{ptW}.  The ${\cal O}(\alpha_s^2)$ QCD corrections in the large transverse momentum region are also known~\cite{keith}.  
However, for many applications these theoretical results are insufficient. With an 
integrated luminosity per experiment at the Tevatron exceeding $1~{\rm fb}^{-1}$ and an expected luminosity at the LHC of about 
$10~{\rm fb}^{-1}/{\rm year}$, the statistical error on $W,Z$ production is becoming smaller  than one percent.
This statistical error sets the scale for the desired theoretical precision, so that the physics potential may be 
fully exploited.  It is then easy to see that for percent
level accuracy, next-to-next-to-leading order (NNLO) QCD computations for electroweak gauge boson production 
are required.  The  available results
for ${\cal O}(\alpha_s^2)$ corrections to inclusive $W,Z$ production~\cite{DY} and to $W,Z$ rapidity distributions~\cite{Anastasiou:2003ds} 
confirm that the NNLO QCD effects are at the level of a few percent, and that the remaining theoretical error after these 
corrections are included is at the percent level or lower.

The experimental identification of the $W$ and $Z$ production 
processes requires cuts on the pseudorapidities and transverse momenta 
of charged leptons and on the missing energy.  The NNLO QCD results to the 
inclusive cross sections and to the rapidity distributions cannot be used to calculate the effects of these cuts, 
since they do not contain the spin correlations between the leptons and the initial-state partons arising from 
the spin-one nature of the electroweak gauge bosons~\cite{Frixione:2004us}.  The fully differential 
${\cal O}(\alpha_s^2)$ corrections with spin correlations included are required to model these effects.  
To illustrate this point further, we discuss three examples.

\begin{enumerate}

\item CDF and D0 have recently presented Run II measurements of the $W/Z$ cross section ratio~\cite{Rmes}. 
With only $72~{\rm pb}^{-1}$ of integrated luminosity, CDF obtained 
\be
R = \frac{\sigma_W \times {\rm Br}(W \to l \nu)}{\sigma_{Z/\gamma^*} 
\times {\rm Br}(Z \to l^+l^-)} = 10.92 \pm 0.15_{\rm stat} \pm 0.14_{\rm sys},
~~~~l = e,\mu.
\ee
The systematic error was estimated to be only 1.5\%.  The actual measurement was performed with the 
cuts $p_T > 20~{\rm GeV}$ and $|\eta| < 1.8$ on the charged lepton, and with missing energy $\not\!\!{E}_T>25$ GeV.  
The results were extrapolated to total cross sections using theoretically computed acceptances. Since no NNLO QCD 
calculation capable of modeling these cuts was available
at the time of that analysis, the procedure for obtaining acceptances was to reweight the PYTHIA rapidity distributions
for the electroweak gauge bosons with the NNLO QCD computation~\cite{Anastasiou:2003ds} to account for NNLO QCD effects. While it is 
unlikely that this procedure leads to drastically wrong results, a precision of a few percent 
{\it cannot be guaranteed} using this technique.

\item The Tevatron and the LHC can potentially provide stringent constraints on parton distribution functions (PDFs).  
The current PDF extractions permit computations of LHC hard-scattering cross sections with $Q \approx 100$ GeV to 
an accuracy of 5\% or better.  Measurements such as the $W,Z$-charge asymmetries~\cite{chargeasym} can reduce 
these errors.  These require precision predictions through NNLO in QCD for leptonic pseudorapidity distributions with 
cuts on transverse momenta and missing energy.

\item With the high luminosity of the LHC, precise measurements 
of electroweak parameters are possible. An interesting 
example is the measurement of the effective electroweak mixing angle 
$\sin^2 \theta_W$ through the forward-backward asymmetry in $Z \to l^+l^-$.  A precision of $2 \times 10^{-4}$, 
competitive with the LEP analysis, can be achieved with $100~{\rm fb}^{-1}$ of 
integrated luminosity, provided the following cuts on the leptons can be imposed:
$|\eta_{e^+,e^-}| < 2.5$, $|Y(e^+,e^-)| > 1$, $p_T^{l} > 20~{\rm GeV}$.
Given the magnitude of the NLO QCD corrections for this set of cuts
\cite{stw}, the inclusion of NNLO QCD effects seems mandatory.

\end{enumerate}

The above discussion illustrates the importance of having a fully differential 
description of electroweak gauge boson production through ${\cal O}(\alpha_s^2)$ in QCD.  
Unfortunately, such computations remain challenging. Their complexity 
lies in the intricate structure of soft and collinear singularities 
that plague individual contributions in QCD perturbation theory. While 
at NLO a number of approaches~\cite{nlogen}  
can be used to isolate and subtract those 
singularities from complicated matrix elements in a process- and observable-independent way, 
an extension of this approach to NNLO is not complete~\cite{nnlosub}.   We have formulated an alternative technique in 
a recent series of papers~\cite{method}.  The central idea 
of this method is an automated extraction of infrared singularities 
from the real radiation matrix elements and a numerical cancellation of 
these divergences with  the virtual corrections.  We have previously applied 
this approach to the computation of the fully differential 
Higgs boson production cross section in gluon fusion  and to $e^+e^- \to~2~{\rm jets}$ through NNLO~\cite{method}.  
We make extensive use of these references in this manuscript.  
Our goal is to describe a fully realistic calculation of single electroweak gauge boson production at the Tevatron and the LHC. 
The computation is valid through NNLO in perturbative QCD, includes spin correlations, finite widths effects, $\gamma-Z$ interference and is fully 
differential. A short version of this paper with initial results was presented in~\cite{wmnnlo}.  

This manuscript is organized as follows. 
In the next Section we briefly recall the 
important features of the method and discuss some of the differences between the current 
calculation and  the fully differential computation of $pp \to H \to \gamma
\gamma$, reported in~\cite{method}. 
In Section~III we demonstrate the possible uses of our numerical program by 
presenting phenomenological results for Tevatron and LHC measurements. 
We conclude in Section~IV.

\section{Details of the computation}

We consider the production of a lepton pair in hadronic collisions,
\be
h_1(P_1) + h_2(P_2) \to V + X \to l_1 + l_2 + X,
\ee
where $V = W,Z$ and $l_{1,2}$ represent charged leptons or neutrinos, as appropriate.  Within the framework of QCD factorization, 
the cross section for this process is 
\be
{\rm d} \sigma^V = \sum_{ij} \int dx_1 dx_2 f_i^{h_1}(x_1) f_j^{h_2}(x_2) 
{\rm d} \sigma_{ij \to V+X}(x_1,x_2),
\ee
where the $f_{i}^h$ are parton distribution functions that 
describe the probability to find a parton $i$ with momentum $xP_h$ in 
the hadron $h$. The partonic cross sections ${\rm d}\sigma_{ij}$ are 
computed perturbatively as an expansion in the strong coupling constant 
$\alpha_s$:
\be
{\rm d}\sigma_{ij} = \sigma^{(0)}_{ij \to V} 
+ \left (\frac{\alpha_s}{\pi} \right) \sigma^{(1)}_{ij \to V} 
+ \left (\frac{\alpha_s}{\pi} \right)^2 \sigma^{(2)}_{ij \to V} + {\cal O}(\alpha_s^3).
\ee
The partonic processes $i  + j \to l_1 + l_2 +  X$ that contribute to electroweak 
gauge boson production differ at each order in the perturbative expansion.
At leading order in $\alpha_s$, only quark-antiquark annihilation channels
contribute, while at NLO the (anti)quark-gluon channels also occur.  At NNLO, the 
gluon-gluon fusion and (anti)quark-(anti)quark scattering channels also contribute.  
We note that all the relevant channels are catalogued in great detail in Ref.~\cite{DY}.

Each partonic process contains several distinct contributions.  We describe these components 
using $Z$-boson production in $q \bar q$ annihilation as an example.  At NNLO, 
this process receives three distinct contributions: (i) the 
two-loop virtual corrections to $q \bar q \to Z$; (ii) the one-loop virtual 
corrections to $q \bar q \to Z + g$; (iii) the tree-level processes $q \bar q \to Z+gg$, $q \bar q \to Z + q\bar q$.  
We refer to these three mechanisms as the double-virtual, the real-virtual, and the double-real emission 
corrections.  When computed separately, these three contributions exhibit numerous soft and collinear singularities.  
To produce a physically meaningful 
result, they should be combined in the presence of an infrared-safe measurement
function. In addition, the collinear renormalization of the 
parton distribution functions $f_i^{h}$ is required.

The double-virtual corrections are the simplest to calculate. They require 
the two-loop massless triangle diagrams that were obtained in~\cite{gonzalves}. If dimensional regularization 
is used to regularize ultraviolet, soft and collinear singularities, 
the result is given by a Laurant series in the regularization parameter 
$\ep = (d-4)/2$, where $d$ is the dimensionality of space-time.  The situation is more complex for the real-virtual and double-real corrections
because they contain real emission matrix elements.  These are finite for non-exceptional final-state momenta, 
but they diverge once an emitted parton
becomes either soft or collinear to another parton. 
The challenge in performing NNLO computations is to extract the divergences from 
the real emission matrix elements 
{\it without} integrating over any kinematic 
parameter that describes the real emission process.  The fully differential nature of the computation remains intact only 
if this can be achieved.

We have developed a technique to accomplish this in a previous series of papers~\cite{method}. 
We describe here the salient features of this method.  Consider a double-real emission contribution to the production of a
$Z$-boson in $q \bar q$ annihilation, 
$q(p_1)+ \bar q (p_2) \to Z(p_Z) + g(p_3)+g(p_4)$. 
We choose a parameterization of the final-state momenta that maps the allowed phase-space  
onto the unit hypercube:
\be
\int {\rm d}^dp_Z {\rm d}^d p_3 {\rm d}^d p_4 \delta^+(p_Z^2 - M_Z^2) 
\delta^+(p_3^2) \delta^+(p_4^2) \delta(p_1+p_2 - p_3 - p_4 - p_Z) 
 = \int \limits_{0}^{1} 
\prod \limits_{i=1}^{5}{\rm d} \lambda_i F(\{\lambda_i \} ).
\ee
The function $F(\{ \lambda_i \})$ depends on the details of
the  parameterization. The invariant 
masses of all particles that participate in the process, such as 
$(p_3+p_4)^2, (p_3+p_Z)^2$, etc., become functions of the parameters
$\lambda_i$.  Soft and collinear singularities in the matrix 
elements occur when some invariant masses reach zero or other 
exceptional values: $(p_3+p_4)^2 \to 0$, $(p_3+p_Z)^2 \to M_Z^2$, etc. Those limits correspond
to the edges of phase-space and generally occur when a subset 
of the $\lambda_i$ approaches zero or unity.  Two things can happen in these limits.  
Preferably, the singular limits occur in a ``factorized'' form, and the singularities can be extracted using 
the simple prescription for plus-distributions:
\be
\lambda^{-1+\ep} = \frac{\delta(\lambda)}{\ep} + \sum \limits_{n=0}^{\infty}
\frac{\ep^n}{n!} \left [\frac{\log (\lambda)}{\lambda} \right ]_+.
\ee
If the singular limits do not factorize but instead appear in an ``entangled'' form, such as 
$1/(\lambda_1 + \lambda_2)$, the singularities
are disentangled using iterated sector decomposition~\cite{sector}. We find that at NNLO all singular limits can be 
reduced to one of these two forms.  When applied to real emission diagrams, this procedure enables 
us to rewrite the real emission contribution as a Laurant series in the regularization parameter $\ep$.
The coefficients of this series can be integrated numerically over phase-space in the presence of 
arbitrary kinematic constraints.  The double-virtual, real-virtual 
and the double-real emission contributions can then be combined and the cancellation of divergences for infrared safe observables can be 
established numerically. A detailed discussion of the method with examples 
of parameterizations used in actual computations can be found 
in Ref.~\cite{method}.  We describe here a few novel aspects of the current calculation.

\begin{itemize}

\item Since electroweak gauge bosons couple to fermions chirally, 
we must specify our treatment of the axial current in $d$-dimensions.
This issue arises from Dirac structures 
of the form ${\rm Tr}_{\rm H}[\Gamma^{(1)} \gamma_5]
{\rm Tr}_{\rm L}[ \Gamma^{(2)} \gamma_5]$, where $\Gamma^{(1,2)}$ denote 
generic products of Dirac matrices and  ${\rm Tr}_{\rm H,L}$
refer to traces over hadronic and leptonic degrees of freedom, respectively.  
Unlike in fully inclusive computations,  these 
traces do not vanish when the final-state phase-space is 
sufficiently constrained.   To deal with these terms we follow the prescription of Ref.~\cite{larin}.
We define the non-singlet axial current by removing $\gamma_5$:
\be
\bar \psi \gamma^\mu \gamma_5 \psi \to \frac{-i \epsilon^{\mu \nu \alpha \beta}}
{3!} \bar \psi \gamma_\nu \gamma_\alpha \gamma_\beta \psi.
\label{axialdef}
\ee
These expressions are equivalent in the $d \to 4$ limit.  Since the Levi-Civita tensor is a four-dimensional object,
it must be combined with the matrix elements only after they are rendered finite.  Although this 
seems to imply that computations should be performed with open Lorentz indices, 
which would be cumbersome in realistic calculations, this can be avoided.  
Since we are interested in the products of two traces which each contains a single $\gamma_5$, 
we only obtain products of two 
Levi-Civita tensors of the form $\epsilon_{\mu_1 \mu_2 \mu_3 \mu_4} \epsilon^{\nu_1 \nu_2 \nu_3 \nu_4}$.  
These can be simplified using the identity 
\be
\epsilon_{\mu_1 \mu_2 \mu_3 \mu_4} \epsilon^{\nu_1 \nu_2 \nu_3 \nu_4} = 
- {\rm det} | g_{\mu}^{\nu}|,~~~~\mu = \mu_1,..\mu_4,~~~\nu = \nu_1...\nu_4,
\ee
which can be easily continued to $d \ne 4$.
We write the  determinant in an explicit form and contract the free indices with the matrix elements. The resulting 
expressions become functions of scalar products of the particle momenta.  The replacement in Eq.~(\ref{axialdef}) 
violates the Ward identity that relates the renormalization of the axial and vector currents, since this definition of $\gamma_5$ does 
not anti-commute with all of the $\gamma_{\mu}$~\cite{larin}.  The following 
additional finite renormalization of the axial contribution must be performed:
\be
Z^{ns}_{5} = 1-\frac{\alpha_s}{\pi}C_F+\frac{\alpha_s^2}{16 \pi^2}\left(22C_F^2-\frac{107}{9}C_FC_A+\frac{2}{9}C_F N_f\right),
\ee
where $N_f$ denotes the number of active fermion flavors and $C_{F,A}$ are the standard QCD Casimir invariants.

\item The differential cross section for the partonic process 
$q \bar q \to Z + X \to e^+e^- + X$ can be written as
\be
|{\cal M}|^2 = \frac{H_{\mu \nu} L^{\mu \nu}}{(q^2 - M_Z^2)^2 + M_Z^2 \Gamma_Z^2},
\ee
where $q^2$ is the invariant mass of the di-lepton pair and $H_{\mu \nu}$ 
and $L_{\mu \nu}$ are the hadronic and leptonic tensors, respectively. 
Since we can define infrared safe observables in QCD using 
leptonic momenta, the lepton tensor 
$L_{\mu \nu}$ is irrelevant for the cancellation of soft and collinear 
singularities.  It is therefore natural to take the leptonic phase-space 
in four dimensions, rather than $d$ dimensions, to simplify the calculation of matrix elements.  
However, we must take care when  writing the scalar products 
of leptonic momenta  with partonic momenta in the double-real emission corrections.  
We consider here the $q\bar{q} \to ggZ \to gge^+e^-$ process for illustration.  
The phase-space for the $gge^+e^-$ final state factorizes into the phase-space for $q\bar q \to ggZ$ and 
the phase-space for $Z \to e^+  e^-$. We perform a Sudakov decomposition of the final-state momenta 
in terms of the incoming partonic momenta $p_1,p_2$.  Denoting the momentum of one of the 
gluons as $p_3$, we have
\be
p_3^\mu = a_3 p_1^\mu +  b_3 p_2^\mu + p_{3T}^\mu.
\ee
Because of momentum  conservation, we must define one relative 
angle between $\vec p_{3 T}$ and $\vec p_{Z T}$ 
to parameterize the $gg Z$ phase-space and 
one relative angle between $\vec p_{e T}$ and $\vec p_{Z T}$ to parameterize 
the $Z \to e^+ e^-$ phase-space.  In $d=4$, each transverse phase-space is 2-dimensional, and 
these two angles determine the relative angle between $\vec p_{3 T}$ and $\vec p_{e T}$. 
In $d \ne 4$, the $ggZ$ transverse phase-space is $(d-2)$-dimensional; an additional angle $\phi$ 
is needed to define the relative orientation between the planes 
defined by $\vec p_{Z T}, \vec p_{3T}$ and by $\vec p_{Z T}, \vec p_{e T}$.  The scalar product must 
therefore  be written as
\be
\vec p_{3 T} \cdot  \vec p_{e T} = 
p_{3 T} p_{e T} \left ( \cos \phi_{3Z} \cos \phi_{eZ} 
 + \sin \phi_{eZ} \sin \phi_{3Z} \cos \phi \right ).
\ee
In the limit $d \to 4$, we have $\cos \phi \to \pm 1$, and this expression reduces to 
$p_{3 T} p_{e T}\cos \left(\phi_{3Z} \pm \phi_{eZ}\right)$.  This makes it explicit that 
given the two angles $\phi_{3Z}$ and $\phi_{eZ}$, the orientation of $\vec p_{3 T}$ and $\vec p_{e T}$ is 
completely determined in $d=4$.

\end{itemize}

After all three components of the hard-scattering cross section are combined, additional
counterterms are needed to remove initial state collinear singularities.  
It is straightforward to extend the numerical approach described in~\cite{method} to obtain the desired results.  

We have essentially two checks on our calculation.  
First, considering  different  cuts on the
leptonic transverse momenta and rapidities
and on the missing energy for $W$ production, we verify
cancellation of the divergences in the production cross sections.
Because the divergences start at $1/\ep^{4}$ at NNLO, the
cancellation of all divergences through $1/\ep$ provides a stringent 
check on the calculation.  We also check that the vector and axial contributions 
are separately finite, as required.  A second check is obtained by 
integrating fully over the final state phase-space and comparing against 
known results for the inclusive cross section.  We find excellent agreement 
with the results of~\cite{DY} for all partonic channels.

\section{Phenomenological results}

We have implemented our calculation into a numerical program, and we now discuss several phenomenological 
results obtained using this code.  We first present the input parameters.
We use the MRST parton distribution functions~\cite{mrst} at the appropriate order 
in $\alpha_s$.  We use $M_Z = 91.1875~{\rm GeV}$, $\Gamma_Z = 2.4952~{\rm GeV}$, 
${\rm Br}(Z \to e^+e^-) = 0.0336$, 
$M_W = 80.451~{\rm GeV}$, $\Gamma_W = 2.118~{\rm GeV}, {\rm Br}(W \to e\nu) = 0.1068$.
We set $|V_{\rm ud}| = 0.974$, $|V_{\rm us}| = |V_{\rm cd}| = 0.219$, 
and $|V_{\rm cs}| = 
0.996$, and obtain $|V_{\rm ub}|$ and $|V_{\rm cb}|$ from unitarity of the CKM matrix.  
We neglect contributions 
from the top quark; these have been shown to be small in the inclusive cross section~\cite{DY}.  
For 
electroweak input parameters, we use $\sin^2 \theta_W = 0.2216$ and  
$\alpha_{\rm QED}(m_Z) = 1/128$. 
We set the factorization and renormalization scales to a common value, $\mu_r=\mu_f=\mu$, and employ 
various choices of $\mu$ in our numerical study.  To perform the numerical integration we use the CUBA package~\cite{Hahn:2004fe}.

\begin{figure}[htb]
\centerline{
\psfig{figure=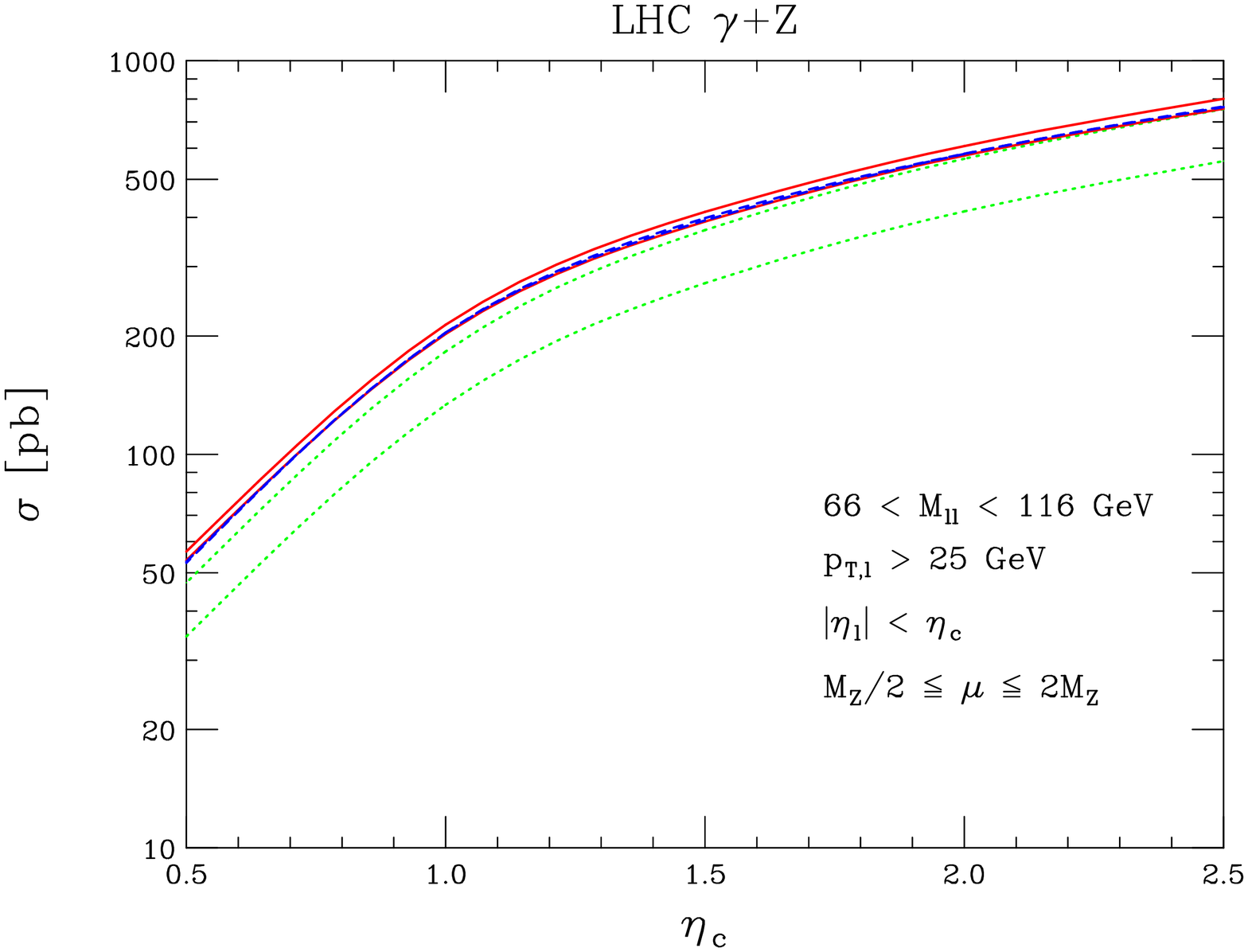,angle=90,width=3in}
\psfig{figure=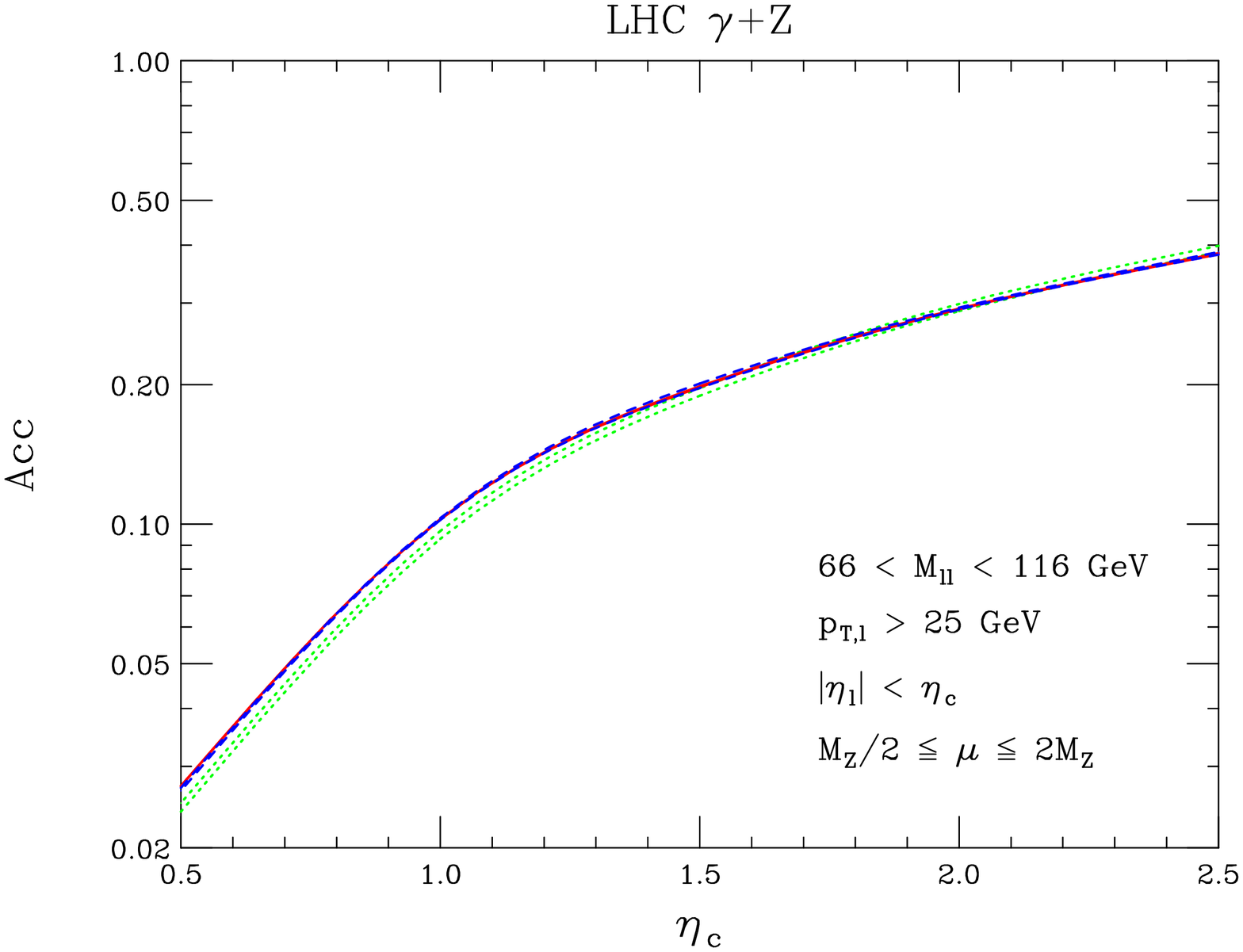,angle=90,width=3in}
}
   \caption{The production cross section (left panel) and 
the acceptance (right panel) as functions of the lepton pseudorapidity cut for neutral current $l^+l^-$ production 
at the LHC.  The two charged leptons are required to have $p_T > 25~{\rm GeV}$, 
and their invariant mass is constrained to $66 < M_{l^+l^-} < 116~{\rm GeV}$.
The dotted green lines refer to the LO result for $\mu=M_Z/2$ and $\mu=2M_Z$, 
the solid red lines indicate the NLO result, and the dashed blue lines denote 
the NNLO result.  We note that the $\mu=M_Z/2$ and $\mu=2M_Z$ NNLO lines almost completely overlap and are
nearly indistinguishable in both panels, 
and that both NNLO lines are completely contained within the NLO results.
\label{fig1}
}
\end{figure}

The identification of electroweak gauge bosons at hadron colliders typically requires cuts on the 
transverse momenta and pseudorapidities of the charged leptons, as well as on the missing energy for $W$-boson production.  
For $Z$ production, the invariant mass of the di-lepton pair 
is also restricted to suppress the importance of photon exchange.   
We first study the importance of the NNLO QCD effects for 
the cross sections and acceptances as a function of kinematic cuts for $Z$ production at the LHC.
In Fig.~\ref{fig1}, the neutral current $l^+l^-$ rate and acceptance at the LHC is studied as a function of a cut on the 
leptonic pseudorapidities.  The NNLO results are absolutely stable with respect to scale variations, 
with residual uncertainties much less than one percent, and are completely contained within the NLO 
uncertainty bands.

\begin{figure}[htb]
\centerline{
\psfig{figure=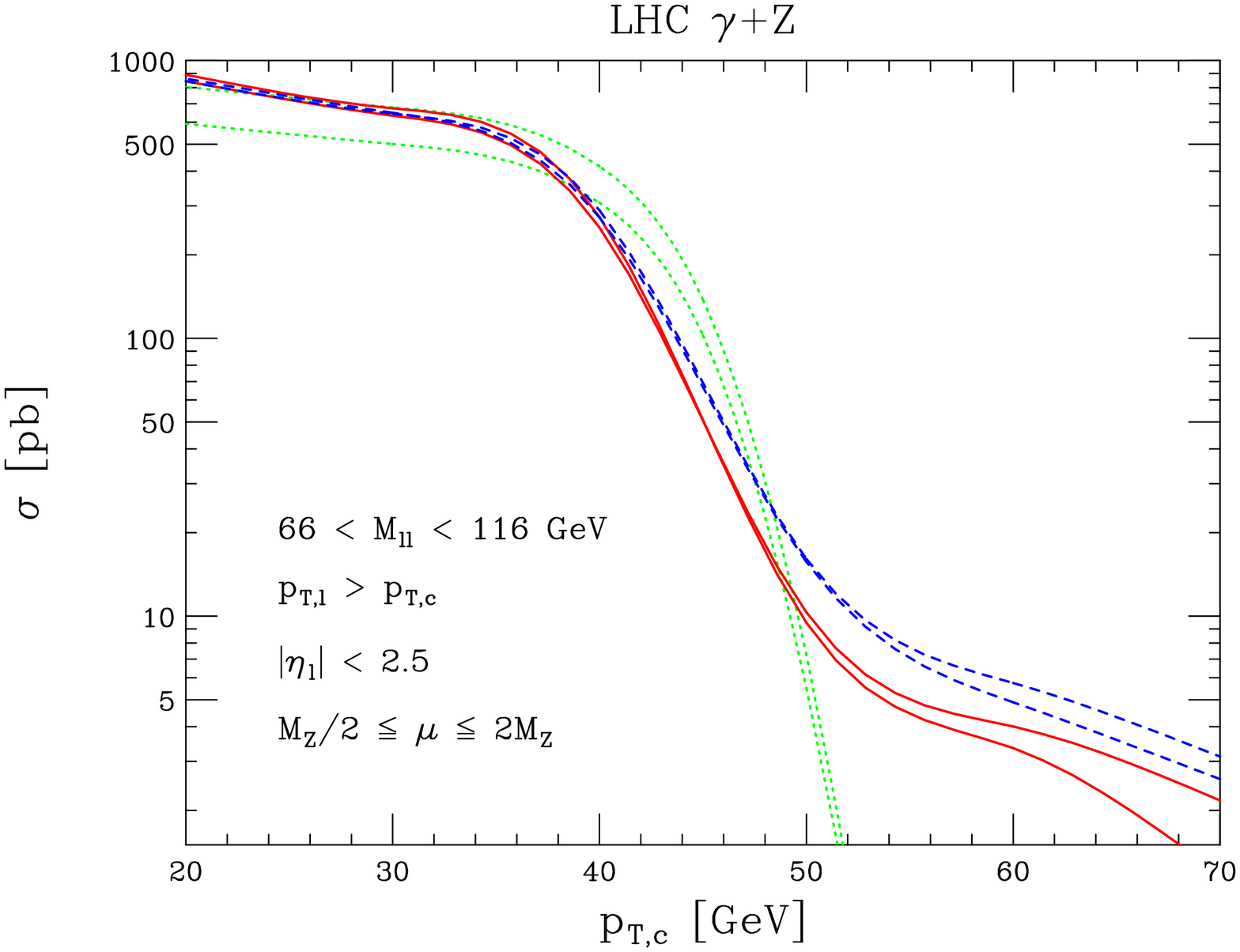,angle=90,width=3in}
\psfig{figure=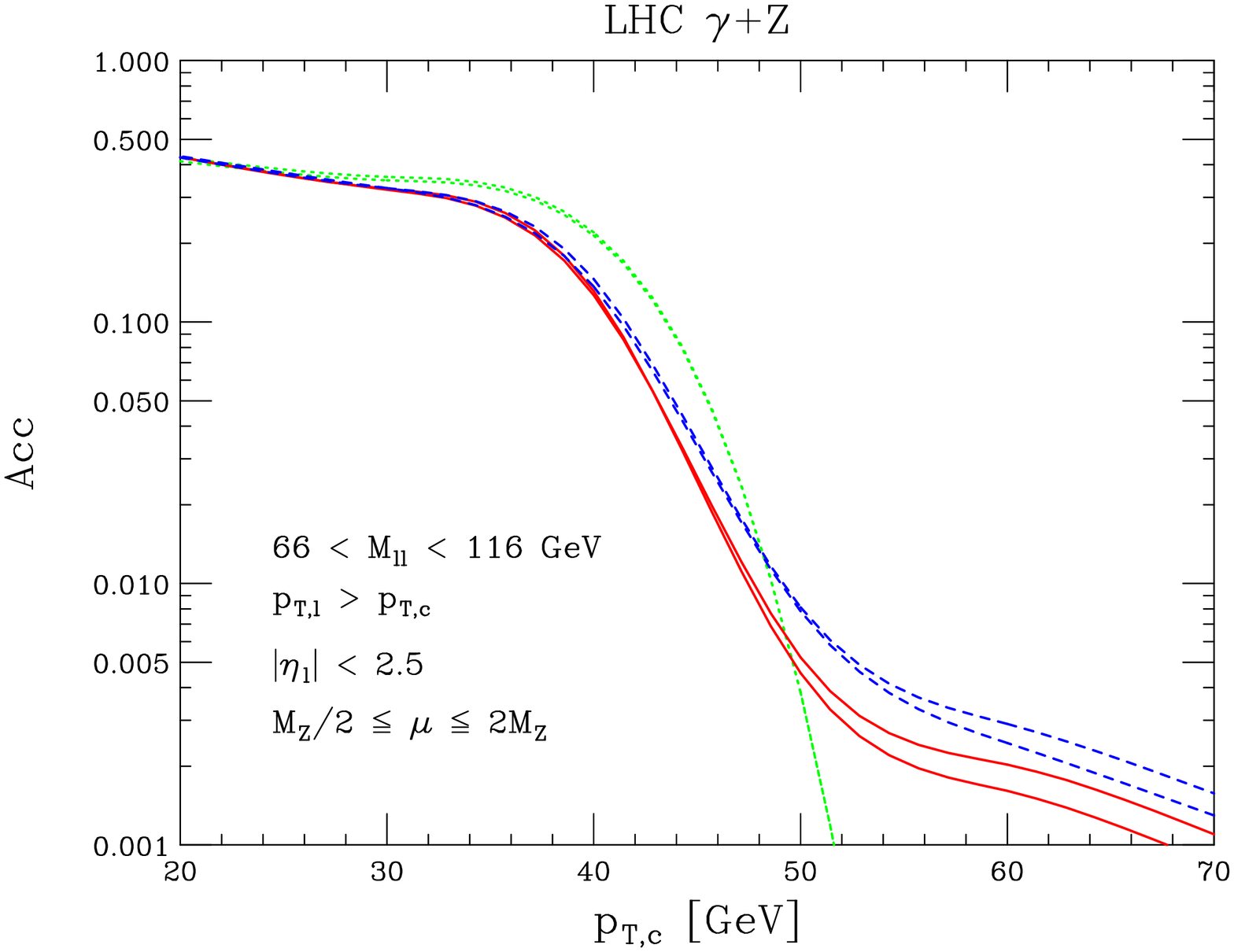,angle=90,width=3in}
}
\caption{The production cross section (left panel) and 
the acceptance (right panel) as functions of the lepton transverse momentum cut for neutral current $l^+l^-$ production 
at the LHC.  The two charged leptons are required to have $|\eta|<2.5$, 
and their invariant mass is constrained to $66 < M_{l^+l^-} < 116~{\rm GeV}$.
The dotted green lines refer to the LO result for $\mu=M_Z/2$ and $\mu=2M_Z$, 
the solid red lines indicate the NLO result, and the dashed blue lines denote 
the NNLO result.
\label{fig2}
}
\end{figure}

In Fig.~\ref{fig2} we present the neutral current $l^+l^-$ rate and acceptance at the LHC as a function of a 
minimum lepton $p_T$ cut which we refer to as $p_{T,c}$.  Several comments regarding these results are required.

\begin{itemize}

\item There is a kinematic boundary at $p^b_T = M_Z/2$ above which the pure $Z$ contribution to the LO cross section vanishes 
in the limit $\Gamma_Z \to 0$.  At higher 
orders, soft gluon effects are important near this boundary.  We expect the fixed-order result to be 
very accurate for values of $p_{T,c}$ away from this boundary.  Evidence for this is provided by the 
close agreement between NLO and MC@NLO for a similar boundary 
at $p^b_T = M_W/2$ in $W$ production~\cite{Frixione:2004us}.

\item Below $p_{T,c}=40$ GeV, the NNLO results are absolutely stable with respect to scale variations, 
with residual uncertainties less than one percent, and are almost completely contained within the NLO 
uncertainty bands.

\item For higher values of $p_{T,c}$, there are large shifts when going from NLO to NNLO, and the scale uncertainties 
underestimate the corrections.  This is not too surprising; in the limit $\Gamma_Z \to 0$ 
the LO result vanishes in this region since an additional radiated 
gluon is needed to have $p_T > M_Z/2$, and what we call NLO is the first term in the perturbative expansion.  The 
absolute magnitude of the shift is also consistent with an ${\cal O}(\alpha_s^2)$ effect.

\end{itemize}

We now discuss a Tevatron analysis of the $W$-boson cross section.  CDF recently presented a measurement 
of the $W \to e\nu$ cross section in the forward rapidity region, $1.2<|\eta|<2.8$, and 
compared this result to the central cross section~\cite{forward}.  
Different values of Bjorken-$x$ contribute to each rapidity region, and 
measuring the central/forward cross section ratio may provide a useful constraint on parton 
distribution functions.  The geometric and kinematic cuts in each region on the charged lepton pseudorapidity and 
transverse momentum, and on the missing energy, are listed below.

\begin{itemize}
\item Forward: $1.2<|\eta|<2.8$, $E_T>20$ GeV, $\not\!\!{E}_T>25$ GeV;
\item Central: $|\eta|<1.1$, $E_T>25$ GeV, $\not\!\!{E}_T>25$ GeV.
\end{itemize}

\noindent In the central cross section analysis there are additional selection cuts requiring the electron to 
be in the fiducial region of the calorimeter, and for the tracker to find an 
electron with $p_T>10$ GeV consistent with the energy deposition in the calorimeter~\cite{Rmes}.  
These cuts give an additional factor $A_{cor} = 0.6985$, so that the acceptance in the 
central region is $A_{cen}=A_{geom} \times A_{cor}$.  We compute $A_{geom}$ through NNLO in perturbative QCD and 
use the given $A_{cor}$ to determine the acceptance in the central region.

We present in Table~\ref{table1} the predictions for the acceptances 
in the central and forward regions, and for the central/forward ratio $R_{c/f}$.  The uncertainties in this table have been obtained by 
computing the results for the scale choices $M_W/2 \leq \mu \leq 2M_W$ and equating the spread with the residual uncertainty.  
This procedure is supported by the NNLO results lying within the ranges indicated by the NLO scale variation.  The NNLO 
theoretical uncertainties are at the $0.25\%$ level or less, and are completely negligible.  The forward region acceptance in 
particular is absolutely stable against radiative corrections.  In the central region we observe an error reduction of a 
factor of five when the NNLO QCD effects are included.  Only experimental errors and parton distribution 
function uncertainties remain, indicating that this measurement can potentially provide useful constraints on 
parton distribution functions.  Our result $R^{th}_{c/f}=0.9266(19)$ is in good agreement with the preliminary value 
obtained by CDF, $R^{exp}_{c/f}=0.925(33)$~\cite{forward}.

\begin{table}[htbp]
\begin{center}
\begin{tabular}{|l|l|l|}
\hline\hline
& NLO & NNLO \\ \hline\hline
$A_{for}$ & $0.2616(2)$ & $0.2614(2)$ \\ \hline
$A_{cen}$ & $0.2458(28)$ & $0.2422(5)$ \\ \hline
$R_{c/f}$ & $0.940(12)$ & $0.9266(19)$ \\ \hline\hline
\end{tabular}
\caption{\label{table1} The theoretical predictions for the forward region acceptance $A_{for}$, the central region 
acceptance $A_{cen}$, and the ratio of central/forward acceptances $R_{c/f}$, together 
with their associated uncertainties at NLO and NNLO.  We note that both the geometric acceptance and 
the factor $A_{cor}$ have been included in the central region result.}
\vspace{-0.1cm}
\end{center}
\end{table}

\section{Conclusions}

In this paper we described a computation of the fully differential cross section through 
NNLO in QCD for $W$ and $Z$ boson production in hadronic collisions.  Our result 
includes spin correlations, finite width effects, $\gamma - Z$ interference and allows for 
the application of arbitrary cuts on the final-state decay products.  We have incorporated 
our result into a numerical code {\small\sf FEWZ} available at the web site {\small\sf http://www.physics.hawaii.edu/$\sim$kirill/FEHiP.htm}.  
We believe this program will be invaluable for precision electroweak studies at 
both the Tevatron and the LHC.

We studied several LHC and Tevatron examples where precise predictions for gauge boson acceptances are 
required.  The theoretical prediction for neutral current $l^+l^-$ production at the LHC is absolutely 
stable with respect to residual scale dependence, with a remaining theoretical uncertainty much less than 
1\%, as long as the minimum $p_T$ cut on the leptons is less than the kinematic boundary value $p^b_T=M_Z/2$.  
For momenta above $p^b_T$ the LO result vanishes in the limit $\Gamma_Z \to 0$.  The NNLO calculation provides the first radiative correction 
in this region, and a significant scale variation remains.

We also studied the $W$ cross section in the central and forward regions, recently analyzed by the 
CDF collaboration.  The theoretical predictions for the acceptances in each region have residual 
uncertainties less than $0.25\%$.  Our calculation of the ratio of central and forward cross sections, 
$R^{th}_{c/f}=0.9266(19)$, is in good agreement with the preliminary CDF result $R^{exp}_{c/f}=0.925(33)$.

\medskip

{\bf Acknowledgments}: 
We thank M. Herndon and especially G. Chiarelli for several helpful discussions 
on the CDF analysis of the $W$ cross section.  
K.M. is supported in part by the DOE grant DE-FG03-94ER-40833, Outstanding 
Junior Investigator Award and by the Alfred P.~Sloan Foundation. 
F.P. is supported in part 
by the University of Wisconsin Research Committee 
with funds provided by the Wisconsin Alumni Foundation, and 
by the Alfred P.~Sloan Foundation. 



\begin{thebibliography}{29}


\bibitem{wmeas}
For reviews, see 
P.~M.~Nadolsky,
  AIP Conf.\ Proc.\  {\bf 753}, 158 (2005)
  [arXiv:hep-ph/0412146];
U.~Baur,
  arXiv:hep-ph/0511064.


\bibitem{lum}
M.~Dittmar, F.~Pauss and D.~Zurcher,
  Phys.\ Rev.\ D {\bf 56}, 7284 (1997)
  [arXiv:hep-ex/9705004];
V.~A.~Khoze, A.~D.~Martin, R.~Orava and M.~G.~Ryskin,
  Eur.\ Phys.\ J.\ C {\bf 19}, 313 (2001)
  [arXiv:hep-ph/0010163];
W.~T.~Giele and S.~A.~Keller,
  arXiv:hep-ph/0104053.


\bibitem{gianotti} 
For a summary of these issues, see F.~Gianotti and M.~L.~Mangano,
  arXiv:hep-ph/0504221.


\bibitem{nlo} 
G.~Altarelli, R.~K.~Ellis and G.~Martinelli,
  Nucl.\ Phys.\ B {\bf 157}, 461 (1979);
J.~Kubar-Andre and F.~E.~Paige,
  Phys.\ Rev.\ D {\bf 19}, 221 (1979);
K.~Harada, T.~Kaneko and N.~Sakai,
  Nucl.\ Phys.\ B {\bf 155}, 169 (1979)
  [Erratum-ibid.\ B {\bf 165}, 545 (1980)];
P.~Aurenche and J.~Lindfors,
  Nucl.\ Phys.\ B {\bf 185}, 274 (1981).



\bibitem{ew1} 
S.~Dittmaier and M.~Kramer,
  Phys.\ Rev.\ D {\bf 65}, 073007 (2002)
  [arXiv:hep-ph/0109062];
U.~Baur, S.~Keller and D.~Wackeroth,
  Phys.\ Rev.\ D {\bf 59}, 013002 (1999)
  [arXiv:hep-ph/9807417].


\bibitem{webber} 
S.~Frixione and B.~R.~Webber,
  JHEP {\bf 0206}, 029 (2002)
  [arXiv:hep-ph/0204244];
S.~Frixione, P.~Nason and B.~R.~Webber,
  JHEP {\bf 0308}, 007 (2003)
  [arXiv:hep-ph/0305252].


\bibitem{ptW} 
C.~Balazs and C.~P.~Yuan,
  Phys.\ Rev.\ D {\bf 56}, 5558 (1997)
  [arXiv:hep-ph/9704258];
F.~Landry, R.~Brock, P.~M.~Nadolsky and C.~P.~Yuan,
  Phys.\ Rev.\ D {\bf 67}, 073016 (2003)
  [arXiv:hep-ph/0212159].


\bibitem{keith} 
R.~K.~Ellis, G.~Martinelli and R.~Petronzio,
  Nucl.\ Phys.\ B {\bf 211}, 106 (1983);
R.~J.~Gonsalves, J.~Pawlowski and C.~F.~Wai,
  Phys.\ Rev.\ D {\bf 40}, 2245 (1989);
P.~B.~Arnold and M.~H.~Reno,
  Nucl.\ Phys.\ B {\bf 319}, 37 (1989)
  [Erratum-ibid.\ B {\bf 330}, 284 (1990)].



\bibitem{DY}
R.~Hamberg, W.~L.~van Neerven and T.~Matsuura,
  Nucl.\ Phys.\ B {\bf 359}, 343 (1991)
  [Erratum-ibid.\ B {\bf 644}, 403 (2002)];
R.~V.~Harlander and W.~B.~Kilgore,
  Phys.\ Rev.\ Lett.\  {\bf 88}, 201801 (2002)
  [arXiv:hep-ph/0201206].

\bibitem{Anastasiou:2003ds}
C.~Anastasiou, L.~J.~Dixon, K.~Melnikov and F.~Petriello,
  Phys.\ Rev.\ Lett.\  {\bf 91}, 182002 (2003)
  [arXiv:hep-ph/0306192];
C.~Anastasiou, L.~J.~Dixon, K.~Melnikov and F.~Petriello,
  Phys.\ Rev.\ D {\bf 69}, 094008 (2004)
  [arXiv:hep-ph/0312266].



\bibitem{Frixione:2004us}
  S.~Frixione and M.~L.~Mangano,
  JHEP {\bf 0405}, 056 (2004) [arXiv:hep-ph/0405130].


\bibitem{Rmes}
D.~Acosta {\it et al.}  [CDF II Collaboration],
  Phys.\ Rev.\ Lett.\  {\bf 94}, 091803 (2005)
  [arXiv:hep-ex/0406078];
A.~Abulencia {\it et al.}  [CDF Collaboration],
  arXiv:hep-ex/0508029;
P.~Petrov  [CDF and D0 Collaborations],
  AIP Conf.\ Proc.\  {\bf 815}, 49 (2006).




\bibitem{chargeasym} 
D.~Acosta {\it et al.}  [CDF Collaboration],
  Phys.\ Rev.\ D {\bf 71}, 052002 (2005)
  [arXiv:hep-ex/0411059];
D.~Acosta {\it et al.}  [CDF Collaboration],
  Phys.\ Rev.\ D {\bf 71}, 051104 (2005)
  [arXiv:hep-ex/0501023].

\bibitem{stw} See, for example, 
S.~Haywood {\it et al.},
  arXiv:hep-ph/0003275.


\bibitem{nlogen}
R.~K.~Ellis, D.~A.~Ross and A.~E.~Terrano,
  Nucl.\ Phys.\ B {\bf 178}, 421 (1981);
W.~T.~Giele and E.~W.~N.~Glover,
  Phys.\ Rev.\ D {\bf 46}, 1980 (1992);
Z.~Kunszt and D.~E.~Soper,
  Phys.\ Rev.\ D {\bf 46}, 192 (1992);
S.~Frixione, Z.~Kunszt and A.~Signer,
  Nucl.\ Phys.\ B {\bf 467}, 399 (1996)
  [arXiv:hep-ph/9512328];
S.~Catani and M.~H.~Seymour,
  Nucl.\ Phys.\ B {\bf 485}, 291 (1997)
  [Erratum-ibid.\ B {\bf 510}, 503 (1997)]
  [arXiv:hep-ph/9605323].


\bibitem{nnlosub} 
S.~Weinzierl,
  JHEP {\bf 0303}, 062 (2003)
  [arXiv:hep-ph/0302180];
W.~B.~Kilgore,
  Phys.\ Rev.\ D {\bf 70}, 031501 (2004)
  [arXiv:hep-ph/0403128];
S.~Frixione and M.~Grazzini,
  JHEP {\bf 0506}, 010 (2005)
  [arXiv:hep-ph/0411399];
G.~Somogyi, Z.~Trocsanyi and V.~Del Duca,
  JHEP {\bf 0506}, 024 (2005)
  [arXiv:hep-ph/0502226];
A.~Gehrmann-De Ridder, T.~Gehrmann and E.~W.~N.~Glover,
  JHEP {\bf 0509}, 056 (2005)
  [arXiv:hep-ph/0505111].

\bibitem{method} 
C.~Anastasiou, K.~Melnikov and F.~Petriello,
  Phys.\ Rev.\ D {\bf 69}, 076010 (2004)
  [arXiv:hep-ph/0311311];
C.~Anastasiou, K.~Melnikov and F.~Petriello,
  Phys.\ Rev.\ Lett.\  {\bf 93}, 032002 (2004)
  [arXiv:hep-ph/0402280];
C.~Anastasiou, K.~Melnikov and F.~Petriello,
  Phys.\ Rev.\ Lett.\  {\bf 93}, 262002 (2004)
  [arXiv:hep-ph/0409088];
C.~Anastasiou, K.~Melnikov and F.~Petriello,
  Nucl.\ Phys.\ B {\bf 724}, 197 (2005)
  [arXiv:hep-ph/0501130].

\bibitem{wmnnlo} 
K.~Melnikov and F.~Petriello,
  Phys.\ Rev.\ Lett.\  {\bf 96}, 231803 (2006)
  [arXiv:hep-ph/0603182].


\bibitem{gonzalves}  
R.~J.~Gonsalves,
  Phys.\ Rev.\ D {\bf 28}, 1542 (1983);
G.~Kramer and B.~Lampe,
  Z.\ Phys.\ C {\bf 34}, 497 (1987)
  [Erratum-ibid.\ C {\bf 42}, 504 (1989)].


\bibitem{sector}
T.~Binoth and G.~Heinrich,
  Nucl.\ Phys.\ B {\bf 585}, 741 (2000)
  [arXiv:hep-ph/0004013];
see also K.~Hepp,
  Commun.\ Math.\ Phys.\  {\bf 2}, 301 (1966);
M.~Roth and A.~Denner,
  Nucl.\ Phys.\ B {\bf 479}, 495 (1996)
  [arXiv:hep-ph/9605420].

\bibitem{larin}
S.~A.~Larin,
  Phys.\ Lett.\ B {\bf 303}, 113 (1993)
  [arXiv:hep-ph/9302240].

\bibitem{mrst} 
A.~D.~Martin, R.~G.~Roberts, W.~J.~Stirling and R.~S.~Thorne,
  Phys.\ Lett.\ B {\bf 531}, 216 (2002)
  [arXiv:hep-ph/0201127].

\bibitem{Hahn:2004fe}
  T.~Hahn,
  Comput.\ Phys.\ Commun.\  {\bf 168}, 78 (2005)
  [arXiv:hep-ph/0404043].

\bibitem{forward}
CDF Collaboration, CDF-note 8119; M. Lancaster, talk given at the 
33rd International Conference On High Energy Physics, 26 Jul - 2 Aug 2006, Moscow.

\end{thebibliography}
\end{document}